\documentclass[12pt]{iopart}
\usepackage{amsthm}
\usepackage{graphicx}
\font\bb=msbm10 at 12pt
\newcommand{\mathbb}[1]{\hbox{\bb #1}} %realna, prirozena cisla, atd.

\newtheorem{claim}{Claim}[section]
\newtheorem{theorem}[claim]{Theorem}
\newtheorem{proposition}[claim]{Proposition}
\newtheorem{lemma}[claim]{Lemma}
\newtheorem{remark}[claim]{Remark}
\newtheorem{example}[claim]{Example}
\begin{document}

\title[Non-Weyl asymptotics for quantum graphs with general coupling conditions]
{Non-Weyl asymptotics for quantum graphs\\ with general coupling conditions}

\author{E. Brian Davies}
\address{Department of Mathematics, King's College London,\\ Strand, London
WC2R 2LS, U.K.} \ead{E.Brian.Davies@kcl.ac.uk}

\author{Pavel Exner}
\address{Doppler Institute for Mathematical Physics and Applied
Mathematics, \\ Czech Technical University, B\v{r}ehov\'{a} 7,
11519 Prague, \\ and  Nuclear Physics Institute ASCR, 25068
\v{R}e\v{z} near Prague, Czechia} \ead{exner@ujf.cas.cz}

%    Information for third author
\author{Ji\v{r}\'{\i} Lipovsk\'{y}}
\address{Institute of Theoretical Physics, Faculty of Mathematics and
Physics, \\ Charles University, V Hole\v{s}ovi\v{c}k\'ach 2, 18000
Prague, \\ and  Nuclear Physics Institute ASCR, 25068
\v{R}e\v{z} near Prague, Czechia} \ead{lipovsky@ujf.cas.cz}

\begin{abstract}
Inspired by a recent result of Davies and Pushnitski, we study
resonance asymptotics of quantum graphs with general coupling
conditions at the vertices. We derive a criterion for the
asymptotics to be of a non-Weyl character. We show that for
balanced vertices with permutation-invariant couplings the
asymptotics is non-Weyl only in case of Kirchhoff or
anti-Kirchhoff conditions. For graphs without permutation
numerous examples of non-Weyl behaviour can be constructed.
Furthermore, we present an insight into what
makes the Kirchhoff/anti-Kirchhoff coupling particular from the
resonance point of view. Finally, we demonstrate a generalization
to quantum graphs with unequal edge weights.
\end{abstract}

%%%%%%%%%%%%%%%%%%%%%%%%%%%%%%%%%%%%%%%%%%%%%%%%%%%%%%%%%%
%%  INTRODUCTION                                        %%
%%%%%%%%%%%%%%%%%%%%%%%%%%%%%%%%%%%%%%%%%%%%%%%%%%%%%%%%%%
\section{Introduction}

Quantum graphs are objects of intense interest -- we refer to
\cite{AGA} for the history of the subject and a rich bibliography.
One of their attractive features is that they are mathematically
accessible while at the same time allowing an investigation of
various effects uncommon in the usual quantum mechanics in
Euclidean space. A very recent example was presented by a paper by
one of us with Pushnitski \cite{DP} in which the high-energy
behaviour of resonances in quantum graphs was investigated and
situations were demonstrated  in which the asymptotics deviates
from the classical Weyl type.

To be specific, the result in \cite{DP} was that in some
situations the graph can have asymptotically \emph{fewer}
resonances than expected. This occurs if the vertex coupling is
the simplest non-trivial coupling possible, usually called
Kirchhoff, and at least one graph vertex is {\em balanced} in the
sense that it is joined to an equal number of finite internal
edges and semi-infinite external edges, which we call leads. Our
goal in this paper is to determine when non-Weyl resonance
asymptotics occurs for quantum graphs with more general vertex
couplings.

By combining some observations from a previous paper of two of us
\cite{EL} with classical results about zeros of exponential sums
\cite{La}, we derive a criterion for a quantum graph to have a
family of resonances with non-Weyl behaviour; cf.
Theorem~\ref{thmeigenvalue} below. We pay particular attention to
quantum graphs in which the coupling condition at every vertex is
invariant under all permutations of the edges joined to the
vertex. Within this class we show in Theorem~\ref{perm_result}
that there are only two cases yielding non-Weyl asymptotics, the
mentioned result of \cite{DP} being one of them. On the other
hand, if one abandons the permutation symmetry one can produce
numerous examples of graphs with such behaviour, even if none of
their vertices is balanced. We explain how that happens and
illustrate the result with a simple example.

We also present considerations that help to explain why the
Kirchhoff conditions (and their counterpart) are so particular
from the resonance point of view. First we discuss the example of
a loop with two leads attached at the same point through a
$\delta$-coupling; we show that ``one half'' of the resonances
escape to infinity as one approaches the Kirchhoff situation. Then
we add a more general discussion showing, in rough terms, that a
balanced vertex with Kirchhoff conditions allows for a partial
decoupling which effectively diminishes the ``size'' of the graph
which enters the asymptotics.

Finally, we discuss a class of more general graphs, whose edges
have weights, in general unequal ones. In
Theorem~\ref{weightedasympt} we show that the results about
non-Weyl resonance asymptotics can be extended to such systems
provided the notion of a balanced vertex is modified: the issue is
whether a certain linear combination of the weights vanishes at
any vertex.

%%%%%%%%%%%%%%%%%%%%%%%%%%%%%%%%%%%%%%%%%%%%%%%%%%%%%%%%%%
%%  PRELIMINARIES                                       %%
%%%%%%%%%%%%%%%%%%%%%%%%%%%%%%%%%%%%%%%%%%%%%%%%%%%%%%%%%%
\section{Preliminaries}%\label{}

We start by recalling a few notions about quantum graphs and, in
particular, some concepts introduced in \cite{EL} and \cite{DP}.
Consider a metric graph $\Gamma$ consisting of a set of vertices
$\mathcal{X}_j$, $\,N$ internal edges with lengths $l_j$, $j =
1,\dots,N$, and $M$ external semi-infinite edges. We equip it with
the Hamiltonian acting on each edge as
$-\mathrm{d}^2/\mathrm{d}x^2$ with appropriate coupling conditions
at the vertices: the domain of the operator consists of all
functions from $W^{2,2}(\Gamma)$ which satisfy
 % ------------- %
 \begin{equation}
   (U_j-I)\Psi_j +i (U_j+I)\Psi_j' = 0\,, \label{coupling}
 \end{equation}
 % ------------- %
at the vertices of the graph, where $U_j$ is a unitary
$\mathrm{deg\,}\mathcal{X}_j \times \mathrm{deg\,}\mathcal{X}_j$
matrix, $I$ stands for the unit matrix and $\Psi_j$ and $\Psi_j'$
are vectors of the values and the outward derivatives of the
relevant functions at the given vertex, respectively. The
condition (\ref{coupling}) is the standard unique
version\footnote{For alternative descriptions of vertex couplings
see \cite{Ku} and \cite{CET}.} of the general coupling description
\cite{KS1}, which was first proposed in \cite{Ha,
KS2}\footnote{Note that the condition was already known in the
general theory of boundary value problems \cite{GG}.} and was
derived for graphs by a straightforward method in \cite{CE}.

Following \cite{EL}, each graph with finitely many edges can be
equivalently treated as a one-vertex-graph with the coupling given
by a single ``large'' unitary matrix $U$; the leads stem from the
single vertex and all the internal edges begin and end at it. The
topology of the original graph is encoded, of course, in the
matrix $U$; more explicitly, the condition $(U-I)\Psi +i
(U+I)\Psi' = 0$ does not couple a pair of edges if they do not
have any end in common.

The entities of interest are resonances of the described
operator, which we will denote by $H_U$. They are conventionally
defined as poles of the analytic continuation of $(H_U-I)^{-1}$,
and they coincide with the poles of the on-shell scattering matrix
on the graph $\Gamma$. It was shown in \cite{EL} that the resonance
positions are determined by the condition
 % ------------- %
 \begin{equation}
   F(k):=\mathrm{det}\,\left[(U-I)\,C_1(k)+ik (U+I)\,C_2(k)\right]=0\,, \label{res}
 \end{equation}
 % ------------- %
where $C_1(k)$, $C_2(k)$ are $(2N+M) \times (2N+M)$ matrices of
the form
 % ------------- %
 $$
  C_j (k) = \mathrm{diag\,}(C_j^{(1)}(k),C_j^{(2)}(k),\dots,C_j^{(N)}(k),
i^{j-1} I_{M\times M})\,, \quad j=1,2\,,
 $$
 % ------------- %
where the first $2N$ blocks are given by
% ------------- %
 \begin{eqnarray*}
   C_1^{(j)} (k) =
   \left(\begin{array}{cc}
   0 & 1 \\
   \sin{kl_j} & \cos{kl_j}\end{array}\right)\,,\qquad
   C_2^{(j)} (k) =
   \left(\begin{array}{cc}
   1 & 0  \\
   -\cos{kl_j} & \sin{kl_j}\end{array}\right)\,,
 \end{eqnarray*}
 % ------------- %
respectively, and $I_{M\times M}$ is the $M\times M$ unit matrix.
There is an important convention to make here. Usually one
associates resonances with resolvent poles in the lower complex
halfplane while those on the real axis are \emph{eigenvalues},
typically embedded in the continuous spectrum. As in \cite{DP},
however, the latter will be also \emph{regarded as resonances} in
this paper.

It was also shown in \cite{EL} that the problem can be
reformulated as an investigation of the compact ``internal''
graph, the influence of the leads being taken into account by
replacing the coupling at the ``external'' vertices to which the
leads are attached by an effective, energy-dependent one. In
particular, the resonance condition (\ref{res}) can be then
rewritten as
 % ------------- %
 $$
    \mathrm{det}\,\left[(\tilde U(k)-I)\,\tilde C_1(k)
    +ik (\tilde U(k)+I)\,\tilde C_2(k)\right]=0 \,,%\label{uk}
 $$
 % ------------- %
where the $2N \times 2N$ matrices $\tilde C_i(k)$, $i = 1,2$,
contain only the parts of $C_i(k)$ corresponding to the internal
edges,
$$
\tilde C_i (k)  = \mathrm{diag\,}(C_i^{(1)}(k),
C_i^{(2)}(k),\dots,C_i^{(N)}(k))
$$
and the unitary matrix $U$ is replaced by the effective $2N \times 2N$ coupling matrix
 % ------------- %
 \begin{equation}
   \tilde U(k):= U_1 -(1-k) U_2 [(1-k) U_4 - (k+1) I]^{-1} U_3\,,
   \label{efu}
 \end{equation}
 % ------------- %
where $U_i$ refer to the block decomposition $U =\left(\begin{array}{cc}
U_1&U_2\\U_3&U_4\end{array}\right)$ related to grouping
of the internal and external edges, respectively.

As in \cite{DP}, an external vertex of the graph $\Gamma$ is
called {\em balanced} if it connects the same number of internal
and external edges, otherwise we say it is {\em unbalanced}.

%%%%%%%%%%%%%%%%%%%%%%%%%%%%%%%%%%%%%%%%%%%%%%%%%%%%%%%%%%
%%  MAIN RESULT                                         %%
%%%%%%%%%%%%%%%%%%%%%%%%%%%%%%%%%%%%%%%%%%%%%%%%%%%%%%%%%%
\section{The main result}\label{s:main}

Since we are interested in the high-energy asymptotics of the
resonances, it is convenient to rewrite the condition (\ref{res})
in terms of the exponentials $\mathrm{e}^{ikl_j}$ and
$\mathrm{e}^{-ikl_j}$. For brevity, we denote $e_j^\pm
:=\mathrm{e}^{\pm ikl_j}$ and
\[
e^\pm := \Pi_{j=1}^N e_j^\pm=e^{\pm ikV}
\]
where $V=\sum_{j=1}^nl_j$ is the size of the finite part of the graph. The
condition (\ref{res}) then becomes

 % ------------- %
\begin{eqnarray}
  F(k)&:=& \mathrm{det\,}\bigg\{\frac{1}{2}[(U\!-\!I)+k(U\!+\!I)]E_1(k)
   +\frac{1}{2}[(U\!-\!I)-k(U\!+\!I)]E_2(k)\nonumber \\ && +k(U\!+\!I)E_3 +\,(U\!-\!I)E_4 \nonumber \\ && +[(U\!-\!I)-k(U\!+\!I)]\,\mathrm{diag\,}
   \left(0,\dots,0,I_{M\times M}\right)\bigg\}\nonumber \\ &=&0 \,, \label{rese}
 \end{eqnarray}
 % ------------- %
where $E_i(k) = \mathrm{diag\,}\left(E_{i}^{(1)},E_{i}^{(2)},
\dots,E_{i}^{(N)},0,\dots,0\right)$, $i=1,2,3,4$, consists of $N$
nontrivial $2\times 2$ blocks
 % ------------- %
 $$
   E_1^{(j)}= \left(\begin{array}{cc}0&0\\-i e_j^+&e_j^+\end{array}\right),
   \ E_2^{(j)} = \left(\begin{array}{cc}0&0\\i e_j^-&e_j^-\end{array}\right),
   \ E_3^{(j)} = \left(\begin{array}{cc}i&0\\0&0\end{array}\right),
   \ E_4^{(j)} = \left(\begin{array}{cc}0&1\\0&0\end{array}\right).
 $$
 % ------------- %
and a trivial $M\times M$ part.

The {\em counting function} $N(R,F)$ of any entire function $F(k)$ is defined by
\begin{equation}
N(R,F)=\#\{ k:F(k)=0\mbox{ and } |k|<R\},\label{NRF}
\end{equation}
where the algebraic multiplicities of the zeros are taken into
account. The functions $F(k)$ that arise in the present paper are
more general than those discussed in \cite{DP} and the asymptotics
of their zeros is controlled by the following classical theorem.

 % ------------- %
 \begin{theorem}\label{exppol}
Let $F(k) = \sum_{r = 0}^{n} k^{\nu_r}a_r(k)
\,\mathrm{e}^{ik\sigma_r}$, where $\nu_r\in\mathbb{R}$, $a_r(k)$
are rational functions of the complex variable $k$ with complex
coefficients that do not vanish identically, and $\sigma_r\in
\mathbb{R}$, $\sigma_0 <\sigma_1< \dots < \sigma_n$. Suppose also
that $\nu_r$ are chosen so that $\lim_{k\to\infty} a_r(k)
=\alpha_r$ is finite and non-zero for all $r$. There exists a
compact set $\Omega\subset \mathbb{C}$, real numbers $m_r$ and
positive $K_r$, $\:r =1,\dots,n$, such that the zeros of $F(k)$
outside $\Omega$ lie in one of $n$ logarithmic strips, each one
bounded between the curves $-\mathrm{Im\,}k + m_r \log{|k|} =\pm
K_r$. The counting function behaves in the limit $R\to\infty$ as
  % ------------- %
 $$
   N(R,F) = \frac{\sigma_n -\sigma_0}{\pi} R +\mathcal{O}(1)\,. %\label{efu}
 $$
 % ------------- %
 \end{theorem}
 \begin{proof}
The claim follows, e.g., from Theorem~6 in \cite{La}. First of all, we
note that there are finitely many zeros (counting multiplicities)
in $\Omega$ which naturally do not influence the asymptotics.
Secondly, our $F(k)$ belongs to the class $\sum_{r=0}^n k^{\nu_r}
(\alpha_r + o(1))\mathrm{e}^{ik\sigma_r}$ treated in Section~7 of
\cite{La}. The localization of zeros described in the theorem
follows directly from the conclusions there.

The number of zeros in the circle with centre $0$ and diameter $R$
can be estimated from above by the number of zeros satisfying
$|\mathrm{Re\,}k|\leq  R$ and from below by the number of zeros
satisfying $|\mathrm{Re\,}k| \leq \sqrt{R^2 -(K +m \log{R})^2}$
with $K := \max_{1\le r\le n} K_r$ and $m := \max_{1\le r\le
n}|m_r|$. Since $R-\sqrt{R^2 -(K +m \log{R})^2} \to 0$ as $R\to
\infty$, both pairs of lines asymptotically approach each other.
The number of zeros in the $j$-th logarithmic strip between two
lines parallel to the imaginary axis is given by relation (7) in
\cite{La} with $c_n$ being replaced by the distance of
corresponding $\sigma_j$'s, hence the asymptotic behaviour of the
counting function is given by the above relation.
 \end{proof}
 % ------------- %

To apply this theorem we have to determine the coefficients
of $e^\pm$ in the resonance condition (\ref{rese}). To this aim it is more
convenient to use the formulation with a compact graph and
energy-dependent matrices, since the last term in (\ref{rese})
then disappears and the relation thus becomes
 % ------------- %
  \begin{eqnarray}
  F(k)&:=& \mathrm{det\,}\bigg\{\frac{1}{2}[(\tilde U(k)-I)+k(\tilde U(k)+I)]\tilde E_1(k)
\nonumber\\&&
  +\frac{1}{2}[(\tilde U(k)-I)-k(\tilde U(k)+I)]\tilde E_2(k)\nonumber\\&&
   + k(\tilde U(k)+I)\tilde E_3+(\tilde U(k)-I)\tilde E_4\bigg\}
\nonumber\\ &=&0\,,\label{endep}
 \end{eqnarray}
 % ------------- %
where $\tilde E_j$ (the first two of them being energy-dependent)
are the nontrivial $2N \times 2N$ parts of the
matrices $E_j$ and $I$ denotes the $2N \times 2N$ unit matrix.

 % ------------- %
 \begin{lemma} \label{coeflemma}
The coefficient of $e^\pm$ in (\ref{endep}) is
$\left(\frac{i}{2}\right)^N\mathrm{det\,}[(\tilde U(k)-I)\pm
k(\tilde U(k)+I)]$.
 \end{lemma}
  % ------------- %
 \begin{proof}
Let us prove the claim for the coefficient of $e^+$. First of all,
note that one does not need to consider the term involving $E_2 (k)$,
since all the entries of $E_2$ diminish the imaginary part of the
argument of the exponential. Furthermore, bearing in mind that the
determinant is not changed by a similarity transformation, we
diagonalize the matrix $E_1(k)$ using the block diagonal matrix
$V$ consisting of $2\times 2$ blocks $\left(\begin{array}{cc}0&-i
\\1&1\end{array}\right)$ and look for the coefficient of $e^+$ in
the determinant
  % ------------- %
  \begin{eqnarray*}
   \mathrm{det\,}\bigg\{\frac{1}{2}V^{-1}[(\tilde U(k)-I)+k(\tilde U(k)+I)]V \tilde E_{1V}
\\
   \quad + kV^{-1}(\tilde U(k)+I)V \tilde E_{3V}+V^{-1}(\tilde U(k)-I)V \tilde E_{4V}  \bigg\}\,,
 \end{eqnarray*}
 % ------------- %
where the transformed matrices $\tilde E_{iV} = V^{-1}\tilde E_{i}
V $ are still block diagonal with the blocks
 % ------------- %
 $$
   \tilde E_{1V}^{(j)}= \left(\begin{array}{cc}e^{+}_j&0\\0&0\end{array}\right),\quad
   \tilde E_{3V}^{(j)} = \left(\begin{array}{cc}0&-i\\0&i\end{array}\right),\quad
   \tilde E_{4V}^{(j)} = \left(\begin{array}{cc}-i&-i\\ i&i\end{array}\right).
 $$
 % ------------- %
In order to obtain the maximum possible imaginary part of the exponent one has
to choose the odd column contributions to the determinant only
from the term containing $\tilde E_{1V}$. Hence the odd columns of
$\tilde E_{4V}^{(j)}$ do not influence the coefficient of $e^+$
and our task simplifies to determining the coefficient of $e^+$ in
 % ------------- %
 $$
   \mathrm{det\,}\{[(\tilde U(k)-I)+k(\tilde U(k)+I)]
   (\textstyle{\frac{1}{2}}
   \tilde E_{1V}+\tilde E_{3V})\}
   = \frac{i^N}{2^N}\,\mathrm{det}[(\tilde U(k)-I)+k(\tilde U(k)+I)]\,e^+.
 $$
 % ------------- %
The argument for $e^-$ is similar, one need not consider now the
term with $\tilde E_{1V}$ and chooses $V$ that diagonalizes
$\tilde E_{2V}$.
 \end{proof}
 % ------------- %

Combining Theorem~\ref{exppol} with the previous lemma we arrive
at the conclusion which is useful in determining whether
resonances of a given quantum graph Hamiltonian have Weyl
asymptotics or not.
 % ------------- %
 \begin{theorem}\label{thmeigenvalue}
Let us assume a quantum graph $(\Gamma,H_U)$ corresponding to
$\Gamma$ with finitely many edges and the coupling at vertices
$\mathcal{X}_j$ given by unitary matrices $U_j$. The asymptotics
of the resonance counting function as $R\to \infty$ is of the form
 % ------------- %
 $$
   N(R,F) = \frac{2W}{\pi} R
   +\mathcal{O}(1)\,,
 $$
 % ------------- %
where we call $W$ the effective size of the graph. One always has
$$
0\leq W\leq V:=\sum_{j=1}^{N}l_j.
$$
Moreover $W<V$ (in other words,
the graph is non-Weyl in the terminology of \cite{DP}) if and only
if there exists a vertex where the corresponding energy-dependent
coupling matrix $\tilde U_j(k)$ has an eigenvalue
$(1-k)/(1+k)$ or $(1+k)/(1-k)$ for all $k$.
 \end{theorem}
 \begin{proof}
It follows from Lemma~\ref{coeflemma} and Theorem~\ref{exppol}
that a graph is non-Weyl \emph{iff} the ``overall-vertex'' coupling
matrix $\tilde U(k)$ has eigenvalue $(1-k)/(1+k)$ or
$(1+k)/(1-k)$.

Note that one need not worry about the limiting
assumptions in the theorem; if $a_0(k)$ or $a_n(k)$ are
$\mathcal{O}(k^{-m})$, one can replace $F(k)$ by $\tilde F(k) := k^m
F(k)$, which adds only a resonance of finite multiplicity at $k=0$,
and apply the argument to $\tilde F(k)$. In general $\Gamma$ is
a multivertex graph, of course, but the ``overall'' unitary
coupling matrix $U$ is block diagonal, hence $\tilde U(k)$ also
decouples into effective coupling matrices for particular vertices
and their eigenvalues can be computed separately.
 \end{proof}
 % -------------

%%%%%%%%%%%%%%%%%%%%%%%%%%%%%%%%%%%%%%%%%%%%%%%%%%%%%%%%%%
%%  SYMMETRIC COUPLING                                  %%
%%%%%%%%%%%%%%%%%%%%%%%%%%%%%%%%%%%%%%%%%%%%%%%%%%%%%%%%%%
\section{Permutation-symmetric coupling}\label{symcoup}

In this section we apply the above results to graphs whose
coupling at every vertex is invariant with respect to permutations
of the edges at the vertex. It is easy to see that the coupling is
described by matrices of the form $U_j = a_j J + b_j I$, where
$a_j$, $b_j$ are complex numbers satisfying $|b_j|=1$ and
$|b_j+a_j \mathrm{deg\,} \mathcal{X}_j| =1$; the symbol $J$
denotes the square matrix all of whose entries equal to one and
$I$ stands for the unit matrix.

The class of coupling conditions with permutation symmetry
includes two most important particular cases --- the
$\delta$-conditions with $U_j = \frac{2}{d_j+i\alpha_j}J - I$,
where $d_j$ is the number of edges emanating from the vertex
$\mathcal{X}_j$ and $\alpha_j\in\mathbb{R}$ is the coupling
strength, and the $\delta'_\mathrm{s}$-conditions corresponding to
$U_j = -\frac{2}{d_j-i\beta_j}J + I$ with $\beta_j\in\mathbb{R}$.
The particular case $\alpha_j = 0$ of the $\delta$-coupling is the
so-called Kirchhoff condition --- \emph{free} would be a better
name --- which was discussed in \cite{DP}.

Let us consider a vertex which connects $p$ internal and $q$
external edges. For simplicity, we omit the subscript $j$ labeling
the vertex. On the other hand, to avoid confusion we mark the size
of the matrices $J$ and $I$. As a preliminary, we state the
following lemma without its proof, which is straightforward.
 % ------------- %
 \begin{lemma}\label{symeigval}
The matrix $U = a J_{n\times n}+b I_{n\times n}$ has $n-1$
eigenvalues $b$ and one eigenvalue $na+b$. Its inverse is $U^{-1}
= -\frac{a}{b(an+b)}J_{n\times n}+\frac{1}{b}I_{n\times n}$.
 \end{lemma}
 % ------------- %
Next we give an explicit expression for the effective coupling.
 % ------------- %
 \begin{lemma}\label{symuk}
Suppose that $p$ internal and $q$ external edges are connected
by means of the coupling given by $U = aJ_{(p+q)\times (p+q)} + b
I_{(p+q)\times (p+q)}$. Then the corresponding energy-dependent
coupling matrix is
  % ------------- %
  $$
    \tilde U (k) = \frac{a b (1-k)
    - a (1+k)}{(aq+b)(1-k)-(k+1)}J_{p \times p}+ b I_{p \times p}\,.
  $$
  % ------------- %
 \end{lemma}
 \begin{proof}
The matrix $\tilde U (k)$ is defined by the expression
(\ref{efu}), where $U_1 = aJ_{p\times p} + bI_{p\times p}$, $U_4 =
aJ_{q\times q} + bI_{q\times q}$, and $U_2$ and $U_3$ are
rectangular $p\times q$ and $q\times p$ matrices, respectively,
with all the entries equal to $a$. The needed inverse is easily
computed with the help of the previous lemma.
 \end{proof}
 % -------------

As the main result of this section, we show that in the whole
two-parameter class of permutation-symmetric coupling conditions
there are only two cases which exhibit non-Weyl asymptotics.
Specifically, a quantum graph with $\delta$ or
$\delta_\mathrm{s}'$-conditions can be non-Weyl only if the
corresponding coupling strength is zero.
 % ------------- %
 \begin{theorem} \label{perm_result}
Let $(\Gamma,H_U)$ be a quantum graph with permutation-symmetric
coupling conditions at the vertices, $U_j = a_j J + b_j I$. Then
it has non-Weyl asymptotics if and only if at least one of its
vertices is balanced in the sense that $p = q$, and the coupling
at this vertex is either
\begin{enumerate}
\item[(a)] $f_m = f_n,\quad \forall m,n\leq 2p$, \quad
$\sum_{m=1}^{2p} f_m' = 0$, \quad \textrm{i.e.}$\:\:U =
\frac{1}{p}J_{2p\times 2p}-I_{2p\times 2p}$\,,
\end{enumerate}
or
\begin{enumerate}
\item[(b)] $f_m' = f_n',\quad \forall m,n\leq 2p$, \quad
$\sum_{m=1}^{2p} f_j = 0$, \quad \textrm{i.e.}$\:\:U =
-\frac{1}{p}J_{2p\times 2p}+I_{2p\times 2p}$\,.
\end{enumerate}
 \end{theorem}
 \begin{proof}
Using Lemmata~\ref{symeigval} and \ref{symuk} together with
Theorem~\ref{thmeigenvalue} we have to determine the values of $a$,
$b$, $p$ and $q$ for which the matrix $\tilde U(k)$ has an eigenvalue $(1
\pm k)/(1\mp k)$. Since $|b|=1$, the only possibility is that the
relation
  % ------------- %
  $$
     a p \frac{b (1-k) - (1+k)}{(aq+b)(1-k)-(k+1)}+ b = \frac{1\pm k}{1\mp k}\,.
  $$
  % ------------- %
holds for all $k$. One of the cases yields
  % ------------- %
  $$
     [a (p + q) +b]b (1-k) - (ap+aq+2b)(1+k)= - \frac{(1+k)^2}{1-k}\,,
  $$
  % ------------- %
which cannot be satisfied for any value of the parameters $a$,
$b$, $p$ and $q$. The other case yields
  % ------------- %
  $$
     \{[a (p + q) +b]b + 1\} (1-k) -(1+k) (ap+b)= (aq+b) \frac{(1-k)^2}{1+k}\,.
  $$
  % ------------- %
Since this has to be true for all $k$ we have
  % ------------- %
  \begin{eqnarray*}
     ap+b= 0, \quad aq+b=0 &\quad\Rightarrow\quad & p =q\,,
\\
     \left[a(p+q)+b\right] b+1 = 0 &\quad\Rightarrow\quad & (ap)^2 =
     1\,,
  \end{eqnarray*}
  % ------------- %
and consequently, the graph is non-Weyl \emph{iff} $U =\pm
\frac{1}{p}J_{p\times p}\mp I_{p\times p}$.
 \end{proof}
 % -------------

%%%%%%%%%%%%%%%%%%%%%%%%%%%%%%%%%%%%%%%%%%%%%%%%%%%%%%%%%%
%%  UNBALANCED GRAPHS WITH NON-WEYL ASYMPTOTICS         %%
%%%%%%%%%%%%%%%%%%%%%%%%%%%%%%%%%%%%%%%%%%%%%%%%%%%%%%%%%%
\section{Unbalanced non-Weyl graphs}\label{sec-unbal}

We next describe a method of constructing unbalanced graphs with
non-Weyl asymptotic behaviour. For simplicity, we only treat a
simple example; an extension to more complicated graphs is
straightforward.

The trick we use is based on replacing the coupling matrix $U$ by
$W^{-1} U W$, where $W$ is a block diagonal matrix of the form
$$
     W = \left(\begin{array}{cc}\mathrm{e}^{i\varphi} I_{p\times p}&0\\0&
     W_4\end{array}\right)\, .
$$
and $W_4$ is a unitary $q\times q$ matrix.
 % ------------- %
 % ------------- %
 \begin{lemma}\label{sym1}
The family of resonances of $H_U$ does not change if the original
coupling matrix $U$ is replaced by $W^{-1} U W$.
 \end{lemma}
 \begin{proof}
If $U=\left(\begin{array}{cc}U_1& U_2 \\U_3& U_4\end{array}\right)$ then
$W^{-1} U W= U_W = \left(\begin{array}{cc}U_1&
U_2W_4\mathrm{e}^{-i\varphi}\\W_4^{-1}U_3\mathrm{e}^{i\varphi}&
W_4^{-1}U_4 W_4\end{array}\right)$. The effective energy-dependent
coupling matrix $\tilde U_W(k)$ corresponding to $W^{-1}U W$ as in (\ref{efu}) is given by
  % ------------- %
\begin{eqnarray*}
     \hspace*{-3em}\lefteqn{\tilde U_W(k) = U_1 -(1-k) U_2\mathrm{e}^{-i\varphi} W_4
     \{W_4^{-1}[(1-k) U_4 - (k+1) I]W_4\}^{-1} W_4^{-1}U_3 \mathrm{e}^{i\varphi}}\\
     &= & \tilde U(k)\, .
\end{eqnarray*}

 % ------------- %
In other words, since $W$ commutes with $C_i (k)$ the resonance
condition (\ref{res}) is not affected by the similarity
transformation.
 \end{proof}
 % ------------- %

\begin{figure}
   \begin{center}
     \includegraphics{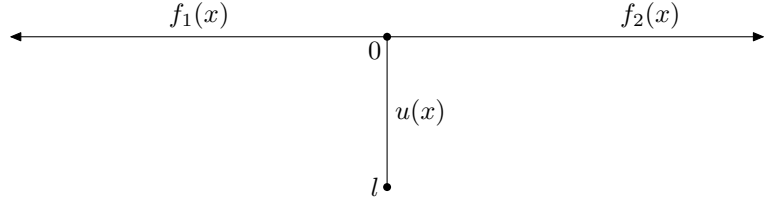}
     \caption{Graph with two leads and one internal edge}\label{fig1}
   \end{center}
\end{figure}

Our next example uses this
transformation to construct unbalanced graphs with non-Weyl asymptotics.
  % ------------- %
 \begin{example}
{\rm  Consider a quantum particle with the Hamiltonian acting as
$-\mathrm{d}^2/\mathrm{d}x^2$ on the graph $\Gamma$ consisting of
two half-lines and one internal edge of length $l$, as sketched in
Figure~\ref{fig1}, studied for the first time in \cite{ES}. The
domain of the Hamiltonian consists of functions from
$W^{2,2}(\Gamma)$ which satisfy the coupling conditions
  % ------------- %
  \begin{eqnarray*}
     0 &=& (U-I)\, ( u(0), f_1 (0),f_2 (0))^\mathrm{T}+ i(U+I)\,(u'(0), f_1' (0),f_2' (0))^\mathrm{T}\,,
\\
     0 &=& u(l) + c u'(l)\,,
  \end{eqnarray*}
 % ------------- %
where $f_i (x)$ are functions on the half-lines and $u(x)$ refers
to the internal edge.

We start from the Robin condition $u(l) + c u'(l) = 0$ at the free
end of the internal edge and the coupling matrix $U_0 =
\left(\begin{array}{ccc}0&1&0\\1&0&0 \\0&0&\mathrm{e}^{i\psi}
\end{array}\right)$.

One may alternatively represent the graph as two disjoint
half-lines; one has a Robin condition parametrized by $\psi$ at
its endpoint, while the other is obtained by joining a half-line
to the original internal edge by means of a free (Kirchhoff)
coupling. Using the original specification, the graph has non-Weyl
asymptotics by \cite{DP} or Theorem~\ref{thmeigenvalue}. Indeed it
cannot have more than two resonances.

We now perform a transformation replacing $U$ by $U_W = W^{-1}U
W$ with
  % ------------- %
  $$
     W  = \left(\begin{array}{ccc}
     1&0&0\\
     0& r \mathrm{e}^{i\varphi_1}&\sqrt{1-r^2}\,\mathrm{e}^{i\varphi_2}\\
     0&\sqrt{1-r^2}\,\mathrm{e}^{i\varphi_3}& -r \mathrm{e}^{i(\varphi_2 +\varphi_3-\varphi_1)}
     \end{array}\right)
 $$
 % ------------- %
and obtain for every fixed value of $\psi$ and $c$ a
three-parameter family of coupling conditions described by the
unitary matrix
  % ------------- %
  $$
     U  = \left(\begin{array}{ccc}
     0& r \mathrm{e}^{i\varphi_1}&\sqrt{1-r^2}\mathrm{e}^{i\varphi_2}\\
     r \mathrm{e}^{-i\varphi_1}& (1-r^2)\mathrm{e}^{i\psi} & -r\sqrt{1-r^2}\mathrm{e}^{-i(-\psi+\varphi_1-\varphi_2)}\\
     \sqrt{1-r^2} \mathrm{e}^{-i\varphi_2}&-r\sqrt{1-r^2}\mathrm{e}^{i(\psi+\varphi_1-\varphi_2)}&r^2\mathrm{e}^{i\psi}
     \end{array}\right)\,,
 $$
 % ------------- %
each of which has the same resonances as $U_0$ by
Lemma~\ref{sym1}. The associated quantum graphs have only two
resonances and are therefore of non-Weyl type. }
 \end{example}

  % ------------- %
 \begin{remark} \label{solom}
{\rm Choosing Dirichlet conditions both at the end of the separated
half-line, $\psi = \pi$, and at the remote end of the internal
edge, $c = 0$, one obtains a family of Hamiltonians which have
no resonances at all. This class contains the example pointed out
in \cite{ES}: the choice $\varphi_1 = \varphi_2 = 0$ and $r =
1/\sqrt{2}$ corresponds to the coupling
 % ------------- %
 $$
     f_1 (0) = f_2(0),\quad  u(0) = \sqrt{2} f_1(0),\quad
     f_1' (0) - f_2'(0)= -\sqrt{2} u'(0)\,,
 $$
 % ------------- %
or in terms of \cite{ES} $b = \sqrt{2}, c = d = 0$. Similarly, the
second possibility indicated in \cite{ES}, i.e. $b = -\sqrt{2}, c
= d = 0$, can be obtained by choosing $\varphi_1 = \varphi_2 = \pi$
and $r = 1/\sqrt{2}$. Notice also that the absence of resonances in
this case is easily understood if one regards the graph in
question as a tree with the root at the remote end of the internal
edge and employs a unitary equivalence proposed first by
Solomyak -- see, e.g., \cite{SS}.}
 \end{remark}
  % ------------- %

%%%%%%%%%%%%%%%%%%%%%%%%%%%%%%%%%%%%%%%%%%%%%%%%%%%%%%%%%%
%%  A LOOP WITH TWO LEADS                               %%
%%%%%%%%%%%%%%%%%%%%%%%%%%%%%%%%%%%%%%%%%%%%%%%%%%%%%%%%%%
\section{A loop with two leads}

\begin{figure}
   \begin{center}
    \includegraphics{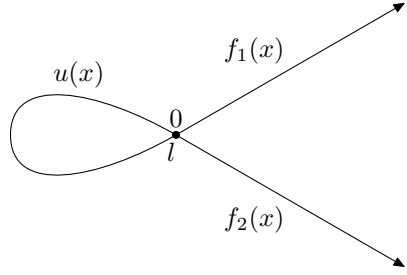}
     \caption{A loop graph with two leads}\label{fig2}
   \end{center}
\end{figure}

We have seen that the resonance asymptotics can be changed by
varying the coupling parameters of the model. This motivates us to analyze
another simple example. The graph $\Gamma$, as sketched on
Figure~\ref{fig2}, now consists of a loop of length $l$ and two
half-lines attached to it at one point. The Hilbert space is
$L^2(0,l)\oplus L^2(\mathbb{R}_+)\oplus L^2(\mathbb{R}_+)$ with
its elements written as $(u(x),f_1(x), f_2(x))^\mathrm{T}$. The
Hamiltonian acts as $-\mathrm{d}^2/\mathrm{d}x^2$ and its domain
will consist of functions from $W^{2,2}(\Gamma)$ satisfying the
conditions
  % ------------- %
  % ------------- %
  % ------------- %
  \begin{eqnarray}
     u(0) = f_1(0)\,,\quad u(l) = f_2(0)\,, \nonumber\\
     \alpha u(0) = u'(0)+ f_1'(0) +\beta (-u'(l)+f_2'(0))\,,\label{coupl}\\
     \alpha u(l) = \beta(u'(0)+ f_1'(0)) -u'(l)+f_2'(0)\nonumber
  \end{eqnarray}
 % ------------- %
with real parameters $\alpha$ and $\beta$. The choice $\beta = 1$
corresponds to the ``overall'' $\delta$-condition of strength
$\alpha$, while $\beta = 0$ decouples two ``inner-outer'' pairs of
meeting edges and one obtains a line with two $\delta$-interactions
at the distance $l$.

In this particular case the resonance condition (\ref{res}) can be
written as
  % ------------- %
  \begin{equation}
    \hspace{-4em} 16\,\frac{-\alpha^2\sin{kl}+2k\alpha(\beta
    + i \sin{kl}-\cos{kl})-2k^2(\sin{kl}
    +i\cos{kl})(\beta^2-1)}{4(\beta^2-1)+\alpha(\alpha-4i)}=0\label{res-two}
  \end{equation}
  % ------------- %

or in terms of $\mathrm{e}^{\pm}$ introduced in
Section~\ref{s:main}
  % ------------- %
  $$
    8 \frac{i\alpha^2 \mathrm{e}^+
    + 4 k \alpha\beta -i[\alpha(\alpha-4ik)
    +4 k^2 (\beta^2-1)]\, \mathrm{e}^-}{4(\beta^2-1)+\alpha(\alpha-4i)}=0\,.
  $$
  % ------------- %
We see that the coefficient of $e^+$ vanishes \emph{iff} $\alpha =
0$, the second term vanishes for $\beta = 0$ or if $|\beta| \not =
1$ and $\alpha = 0$, while the polynomial multiplying $e^-$ does
not vanish for any combination of $\alpha$ and $\beta$.

In other words, the graph has non-Weyl asymptotics \emph{iff}
$\alpha =0$. If, in addition, $|\beta|\not = 1$, then all
resonances are confined to some circle, i.e. the graph has zero
``effective size''. The only exceptional cases are the Kirchhoff
condition, $\beta = 1$ and $\alpha=0$, and its counterpart
condition $\beta = - 1$ and $\alpha=0$, for which one half of the
resonances is asymptotically preserved, in other words, the
effective size of the graph is $l/2$.

Let us look at the $\delta$-condition, $\beta = 1$, in more detail
in order to illustrate the disappearance of half of the resonances
when the coupling strength vanishes. The resonance equation
(\ref{res-two}) becomes
  % ------------- %
  $$
    \frac{-\alpha\sin{kl}+2k(1 + i \sin{kl}-\cos{kl})}{\alpha-4i}=0\,.
  $$
  % ------------- %
A simple calculation shows that the graph Hamiltonian has a
sequence of embedded eigenvalues, $k = 2 n \pi /l$ with
$n\in\mathbb{Z}$, and a family of resonances given by solutions to
$\mathrm{e}^{ikl}=-1+\frac{4ik}{\alpha}$. The former do not depend
on $\alpha$, while the latter behave for small $\alpha$ like
  % ------------- %
  $$
    \mathrm{Im\,}k = -\frac{1}{l} \mathrm{log\,}\frac{1}{\alpha}
    +\mathcal{O}(1)\,, \quad  \mathrm{Re\,}k = n \pi +\mathcal{O}(\alpha)\,,
  $$
  % ------------- %
therefore all the (true) resonances escape to infinity as
$\alpha\to 0$.

%%%%%%%%%%%%%%%%%%%%%%%%%%%%%%%%%%%%%%%%%%%%%%%%%%%%%%%%%%
%%  CAUSES                                              %%
%%%%%%%%%%%%%%%%%%%%%%%%%%%%%%%%%%%%%%%%%%%%%%%%%%%%%%%%%%
\section{What can cause non-Weyl asymptotics}

We have seen that once we allow for general (self-adjoint)
coupling at graph vertices the non-Weyl asymptotic behaviour of
resonances may occur easily. On the other hand, the argument was
based on properties of graphs with edge-permutation symmetries.
The most interesting question posed by the result of \cite{DP} is
thus what makes balanced graphs with Kirchhoff coupling --- or its
``anti-Kirchhoff'' counterpart of Theorem~\ref{perm_result} ---
particular among such graphs.

\subsection{Kirchhoff ``size reduction''}\label{reduction}

\begin{figure}
   \begin{center}
     \includegraphics{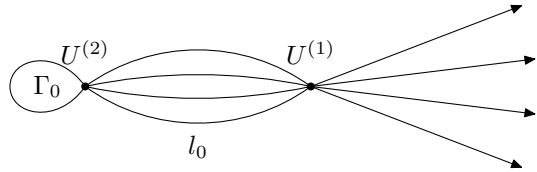}
     \caption{Graph with a balanced vertex considered in Proposition~\ref{transfbal}}\label{fig4}
   \end{center}
\end{figure}

To answer this question, consider
the graph $\Gamma$ sketched in Figure~\ref{fig4}. It contains a
balanced vertex $\mathcal{X}_1$ which connects $p$ internal edges
of the same length $l_0$ and $p$ external edges with the coupling
(\ref{coupling}) given by a unitary matrix $U^{(1)} = a
J_{2p\times 2p} + b I_{2p\times 2p}$. The coupling of the other
internal edge ends to the rest of the graph, denoted as
$\Gamma_0$, is described by a $q\times q$ matrix $U^{(2)}$, where
$q \geq p$; needless to say such a matrix can hide different
topologies of this part of the graph.

Notice first that switching off the coupling between the
``inner-outer'' pairs of edges may lead to different results. If,
for instance, the coupling given by $U^{(1)}$ is Kirchhoff and the
decoupling leads to Kirchhoff at each pair, the size of the graph
is diminished by $pl_0$. If on the other hand, we start from a
nontrivial $\delta$-coupling and the decoupling leads to a
$\delta$-interaction on each edge pair, the graph size remains
preserved. The question is now whether the return to the ``full''
coupling in the former case can restore the size.

To demonstrate that this is not the case and that the effective size
of the graph is smaller in the case of Kirchhoff and
``anti-Kirchhoff'' condition at the balanced vertex one can employ
a trick similar to the one used in Lemma~\ref{sym1}.

  % ------------- %
 \begin{proposition} \label{transfbal}
Let $\Gamma$ be the described graph with the coupling given by
arbitrary $U^{(1)}$ and $U^{(2)}$. Let $V$ be an arbitrary unitary
$p\times p$ matrix, $V^{(1)}:=\mathrm{diag\,}(V,V)$ and
$V^{(2)}:=\mathrm{diag\,}(I_{(q-p)\times(q-p)},V)$ be $2p \times
2p$ and $q\times q$ block diagonal matrices, respectively. Then
$H$ on $\Gamma$ is unitarily equivalent to the Hamiltonian $H_V$
on topologically the same graph with the coupling given by the
matrices $[V^{(1)}]^{-1}U^{(1)}V^{(1)}$ and
$[V^{(2)}]^{-1}U^{(2)}V^{(2)}$.
 \end{proposition}
 \begin{proof}
Let $u$ be an element of the domain of $H$, $u_1,\dots,u_p$ its
restrictions to the internal edges  emanating from
$\mathcal{X}_1$, $f_1,\dots,f_p$ its restrictions to the external
edges  emanating from $\mathcal{X}_1$ and $u_0$ its restriction to
the rest of the graph $\Gamma_0$. Then the map
 % ------------- %
 \begin{eqnarray*}
 (u_1,\dots,u_p)^\mathrm{T} \mapsto (v_1,\dots,v_p)^\mathrm{T} = V^{-1}(u_1,\dots,u_p)^\mathrm{T}
 \\
 (f_1,\dots,f_p)^\mathrm{T} \mapsto (g_1,\dots,g_p)^\mathrm{T} = V^{-1}(f_1,\dots,f_p)^\mathrm{T}
 \\
 u_0(x) \mapsto v_0(x) = u_0(x)
 \end{eqnarray*}
 % ------------- %
is a bijection of the domain of $H$ onto the domain of $H_V$. One
can easily check that the equation (\ref{coupling}) with the
coupling matrices $U^{(1)}$ and $U^{(2)}$ transforms into that
with the coupling matrices $[V^{(1)}]^{-1}U^{(1)}V^{(1)}$ and
$[V^{(2)}]^{-1}U^{(2)}V^{(2)}$.
 \end{proof}
 % ------------- %

  % ------------- %
 \begin{remark}
{\rm The assumption that the internal edges emanating from
$\mathcal{X}_1$ have the same length is made for convenience only.
If it is not satisfied one can still use the above result. To this
aim it is sufficient to denote the length of the shortest of these
edges by $l_0$ and to introduce additional vertices with Kirchhoff
condition at the remaining edges at the distance $l_0$ from
$\mathcal{X}_1$. Similarly, if there is a loop at $\mathcal{X}_1$,
one can introduce a vertex with Kirchhoff condition in the middle
of it.}
 \end{remark}
  % ------------- %

The application of the above result to symmetric couplings, $U^{(1)} =
a J_{2p\times 2p} + b I_{2p\times 2p}$ at $\mathcal{X}_1$, is
straightforward. It is enough to choose the columns of $V$ as an
orthonormal set of eigenvectors of the corresponding $p\times p$
block $aJ_{p\times p} + b I_{p\times p}$, the first one of them
being $\frac {1}{\sqrt{p}}\,(1,1,\dots,1)^\mathrm{T}$. The
transformed matrix $[V^{(1)}]^{-1}U^{(1)}V^{(1)}$ decouples then
into blocks connecting only the pairs $(v_j,g_j)$. The first one
of these, corresponding to a symmetrization of all the $u_j$'s and
$f_j$'s, leads to the $2\times 2$ matrix $U_{2\times 2} = ap
J_{2\times 2} + b I_{2\times 2}$, while the other lead to
separation of the corresponding internal and external edges
described by the Robin conditions  $(b-1)v_j(0) + i (b+1)v_j'(0)
= 0$ and $(b-1)g_j(0) + i (b+1)g_j'(0) = 0$ for $j=2,\dots ,p$.
Notice that this resembles again the reduction of a tree graph due
to Solomyak which we have mentioned in Remark~\ref{solom}.

It is easy to see that the ``overall''
Kirchhoff/anti-Kirchhoff condition at $\mathcal{X}_1$ is
transformed to the ``line'' Kirchhoff/anti-Kirchhoff condition in
the subspace of permutation-symmetric functions. Hence the size
of the graph is reduced by $l_0$ in the Kirchhoff case. The
same is true in the anti-Kirchhoff situation since these
conditions on the line are trivially equivalent to Kirchhoff, in
other words, to the absence of any interaction. In all the other
cases the point interaction corresponding to the matrix $ap
J_{2\times 2} + b I_{2\times 2}$ is nontrivial, and consequently,
the graph size is preserved.

\subsection{Calculating the effective size}\label{calculating}

Since the occurrence of non-Weyl asymptotics for a graph $\Gamma$
subject to Kirchhoff boundary conditions depends on the existence
of a balanced vertex, one might hope to calculate the effective
size of non-Weyl graphs by using some geometrical rules that
quantify the effect of each balanced vertex on the asymptotics.
The following example suggests that this will not be easy. The
effective sizes of the highly symmetrical graphs discussed below
depend on whether a certain integer is or is not equal to $0$ mod
$4$, and calculating the effective sizes is therefore a global
rather than a local issue.

\begin{figure}
   \begin{center}
     \includegraphics{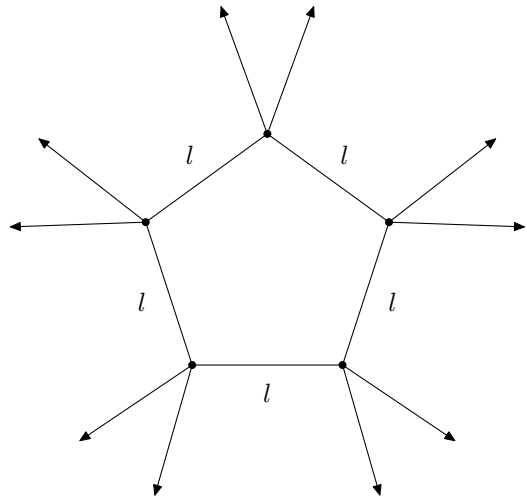}
     \caption{A polygon with pairs of leads considered in Theorem~\ref{polygon-ex}}\label{fig5}
   \end{center}
\end{figure}

We construct a graph $\Gamma_n$ for each integer $n\geq 3$ by
starting with a regular $n$-gon, each edge of which has length
$\ell$. We then attach two semi-infinite leads to each vertex, so
that each of the $n$ vertices is balanced; cf.~Figure~\ref{fig5}.
It follows that the effective size $W_n$ of the graph is strictly
less than the actual size $V_n=n\ell$.

\begin{theorem} \label{polygon-ex}
The effective size of the graph $\Gamma_n$ is given by
$$
W_n=\left\{ \begin{array}{ll}
n\ell/2&\mbox{ if } n\not= 0\,\, \bmod 4,\\
(n-2)\ell/2&\mbox{ if } n= 0\,\, \bmod 4.
\end{array}\right.
$$
\end{theorem}

\begin{proof}
We start by constructing the Bloch/Floquet decomposition of $H$
with respect to the cyclic rotation group $\mathbb{Z}_n$. Let $T$
denote the unitary rotation operator in $L^2(\Gamma_n)$ which
takes each vertex to the next vertex in the clockwise direction.
We observe that $L^2(\Gamma_n)$ is the orthogonal direct sum of
$\mathcal{H}_\omega$ over all complex $\omega$ satisfying
$\omega^n=1$, where
$$
\mathcal{H}_\omega =\{f\in L^2(\Gamma_n):Tf=\omega f\}.
$$
Since the Hamiltonian $H$ commutes with $T$ it follows that the
set of resonances of $H$ is obtained as the union of the
corresponding quantities for the restrictions $H_\omega$ of $H$ to
$\mathcal{H}_\omega$. Our main task is to compute these
explicitly.

\begin{figure}
   \begin{center}
     \includegraphics{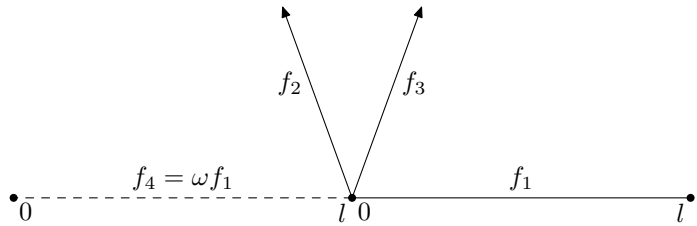}
     \caption{The four-edge graph cell considered in the proof of Theorem~\ref{polygon-ex}}\label{fig6}
   \end{center}
\end{figure}

Let us fix an $\omega$ satisfying $\omega^n=1$. The resonances of
$H_\omega$ can be determined by restricting attention to a cell
consisting of the four edges that are attached to a chosen vertex
-- cf.~Figure~\ref{fig6}. A typical resonance eigenfunction $f$
must be of the form
\begin{eqnarray*}
f_1(x)&=& \alpha \mathrm{e}^{ikx}+\beta \mathrm{e}^{-ikx}, \hspace{2em} 0\leq x\leq \ell,\\
f_2(x)&=& (\alpha+\beta)\mathrm{e}^{ikx}, \hspace{2em} 0\leq x<\infty,\\
f_3(x)&=& (\alpha+\beta)\mathrm{e}^{ikx}, \hspace{2em} 0\leq x<\infty,\\
f_4(x)&=& \omega f_1(x), \hspace{2em} 0\leq x\leq \ell,
\end{eqnarray*}
where the vertex $v$ corresponds to the choice $x=0$ for $f_1,\, f_2,\, f_3$ and to $x=\ell$ for $f_4$.

The continuity condition and Kirchhoff boundary condition at $v$ are
\begin{eqnarray*}
\omega\alpha\mathrm{e}^{ik\ell}+\omega\beta\mathrm{e}^{-ik\ell}%
&=& \alpha+\beta,\\
\omega\alpha\mathrm{e}^{ik\ell}-\omega\beta\mathrm{e}^{-ik\ell}%
&=& 2(\alpha+\beta)+\alpha-\beta.
\end{eqnarray*}
After evaluating the relevant $2\times 2$ determinant one finds
that one has a resonance at $k$ if and only if
\begin{equation} \label{poly-res}
-2(\omega^2+1)+4\omega \mathrm{e}^{-ik\ell}=0.
\end{equation}
Therefore the effective size $W_\omega$ of the system of
resonances of $H_\omega$ is $\ell/2$ if $\omega^2+1\not=0$ but it
is $0$ if $\omega^2+1=0$. Now $\omega^2+1=0$ is not soluble if
$\omega^n=1$ and $n\not=0 \bmod 4$, but it has two solutions if
$n=0 \bmod 4$. The statement of the theorem now follows from the
observation that the effective size $W_n$ of $\Gamma_n$ is given
by $W_n=\sum_{\omega^n=1} W_\omega$.
\end{proof}

\noindent Note that if one puts $\omega=e^{i\theta}$ in
({\ref{poly-res}) then one obtains
$$
k=\frac{1}{l}\Big(i\log(\cos(\theta))+2\pi n\Big)
$$
where $n\in \mathbb{Z}$ is arbitrary. From this equation it is
clear that the resonances all diverge to $\infty$ as
$\theta\to\pm\pi/2$.

%%%%%%%%%%%%%%%%%%%%%%%%%%%%%%%%%%%%%%%%%%%%%%%%%%%%%%%%%%
%%  WEIGHTED GRAPHS                                     %%
%%%%%%%%%%%%%%%%%%%%%%%%%%%%%%%%%%%%%%%%%%%%%%%%%%%%%%%%%%
\section{A generalization: graphs with weighted edges}

In this final section we discuss another
generalization. It is useful to realize that Laplacians on metric
graphs are just the simplest model for networks of quantum wires.
One can, for instance, consider Sturm-Liouville operators on
weighted $L^2$ spaces which describe thin networks of
wires with different, in general varying cross sections \cite{KZ,
EP}. Here we will investigate another possibility which
corresponds to scaling the wires longitudinally. This can have a
physical interpretation; recall that the Hamiltonian on an edge is in
reality $-\frac{\hbar^2}{2m^*} \frac{\mathrm{d}^2}{\mathrm{d}x^2}$
where $m^*$ is the appropriate effective mass, hence joining wires
of different materials we obtain such a scaling.

Consider a general graph $\Gamma$ with
finitely many edges, equipped now with the Hamiltonian
$- c_e^2 \mathrm{d}^2/\mathrm{d}x^2$ on the edge $e$, the
$c_e$'s being positive constants. Its domain consists of functions
from $W^{2,2}(\Gamma)$ satisfying the generalized Kirchhoff
coupling conditions at each vertex $v$, namely
  % ------------- %
  \begin{equation} \label{weight_kirch}
    \sum_{e: v \in e} c_e^2 f_e'(v) = 0\,,\quad f_{e_i}(v)
    = f_{e_j}(v)\quad \mbox{ whenever } v \in e_i,e_j\, . \label{genKirch}
  \end{equation}
  % ------------- %
In other words, the values of the functions are continuous at each
vertex and the sum of the outward derivatives weighted by $c_e^2$
vanishes. For the sake of simplicity we refrain from
discussing the general coupling conditions.

Before turning to the general theory, we illustrate the
asymptotics of weighted graphs by the following simple example.
  % ------------- %
  \begin{example}
\upshape Consider again the graph consisting of one internal edge and
two leads, as in Figure~\ref{fig1}. The Hamiltonian acts now as $-
c^2\, \mathrm{d}^2/\mathrm{d}x^2$ at the internal edge and as $-
\tilde c_j^2\, \mathrm{d}^2/\mathrm{d}x^2$, $j=1,2$ at the leads.
We employ the coupling conditions (\ref{weight_kirch}) at the
junction with the wavefunctions denoted as in Figure~\ref{fig1},
and Dirichlet condition at the free end of the internal edge,
  % ------------- %
  $$
    f_1(0) = f_2(0) = u(0)\,,\quad \tilde c_1^2 f_1(0)'
    + \tilde c_2^2 f_2(0)' +  c^2 u(0)' = 0\,,\quad u(l) = 0\,.
  $$
  % ------------- %
Solving the Schr\"odinger equation explicitly for energy $k^2$ one
gets from the coupling condition at the junction
  % ------------- %
  $$
    c^2\,  \frac{k}{ c} \cos{\frac{kl}{c}}
    = \tilde c_1^2\, \frac{ik}{\tilde c_1} \sin{\frac{kl}{c}}
    + \tilde c_2^2\, \frac{ik}{\tilde c_2} \sin{\frac{kl}{c}}
  $$
  % ------------- %
which leads to
  % ------------- %
  $$
    (c - \tilde c_1 -\tilde c_2) \,\mathrm{e}^{ikl/c}
    + (c + \tilde c_1 + \tilde c_2) \,\mathrm{e}^{-ikl/c} = 0\,.
  $$
  % ------------- %
Consequently, this graph is non-Weyl \emph{iff} $c = \tilde c_1 +
\tilde c_2$, which can be regarded as a generalization of the
Kirchhoff coupling conditions.  The topology of the graph, namely
the fact that the single vertex is unbalanced, has no significance
in this example.
  \end{example}
  % ------------- %

Our general asymptotic theorem for weighted graphs may be proved
by two methods. The calculations in \cite{DP} may be modified line
by line to include the relevant weights on each edge. We describe
here an alternative approach which is simpler in the light of the
analysis in this paper. It depends on the fact that the weighted
graph can be transformed into a graph with all the weights equal
to one by changing the lengths of the edges. This leads to
boundary value conditions at each vertex of a type that were not
considered in \cite{DP}. The interest of our main result,
Theorem~\ref{weightedasympt}, is that it leads to a new insight
into the nature of the `balanced vertex' condition.

  % ------------- %
  \begin{lemma} \label{scale_tr}
Let $\Gamma$ be a weighted graph with the lengths of the internal
edges $l_j$, their weights $c_j^2$, and the weights of the
external edges $\tilde c_j^2$, both entering the coupling
conditions (\ref{genKirch}). Then the corresponding weighted
Hamiltonian on $\Gamma$ is unitarily equivalent to the
non-weighted operator, i.e. the Laplacian, on the graph $\Gamma'$
with the lengths of the edges $l_j/c_j$ and the coupling
  % ------------- %
  $$
    \sum_{e: v \in e} \sqrt{c_e}\, g_e'(v) = 0\,,\quad
    \frac{1}{\sqrt{c_{e_i}}}\, g_{e_i}(v)
    = \frac{1}{\sqrt{c_{e_j}}}\, g_{e_j}(v)\quad \mathrm{ whenever\,\, }
    v \in e_i,e_j
  $$
  % ------------- %
with $c_e$ being the common symbol for both the $c_j$ and $\tilde c_j$.
  \end{lemma}
  \begin{proof}
The transformation of the wavefunction on each internal and
external edge $f_j (x) \mapsto g_j(y) = \sqrt{c_j} f_j(c_j y)$, $y
\in (0,l_j/c_j)$ leads to $g_j(0) =\sqrt{c_j} f_j(0)$ and $
g_j'(0) = \sqrt{c_j^3} f_j'(0)$, and similarly for the external
edges. Substituting these expressions into (\ref{genKirch}) we
arrive at the indicated coupling for the $g_j$'s.
  \end{proof}
  % ------------- %

Notice that in view of the previous lemma the natural definition of
the size of the weighted graph is $V = \sum_j l_j/c_j$ where the
sum runs over all the internal edges. Using the lemma in combination
with the results of Sec.~\ref{reduction} one can prove the following.

  % ------------- %
  \begin{theorem}\label{weightedasympt}
Let $\Gamma$ be a weighted graph with the weights described above
and the coupling conditions (\ref{genKirch}). Then the Hamiltonian
has non-Weyl asymptotics iff for at least one of its vertices $v$
the relation
  % ------------- %
  $$
    \sum_{e:v\in e } c_e = \sum_{\tilde{e}:v\in\tilde{e} } c_{\tilde{e}}
  $$
  % ------------- %
holds, where $e$ and $\tilde{e}$ stand for the internal and external
edges emanating from the vertex $v$, respectively.
  \end{theorem}
  \begin{proof}
With the help of the transformation from Lemma~\ref{scale_tr} one
can use an argument analogous to that of Sec.~\ref{reduction}. For
simplicity, we denote the length of the internal edges of the
rescaled graph emanating from the vertex $\mathcal{X}_1$ by $l_1'
= l_1/c_1,\dots,l_p'=l_p/c_p$ and the weights of the corresponding
internal and external edges by $c_1,\dots,c_p$ and $\tilde
c_1,\dots,\tilde c_q$, respectively. Since nothing prevents us
from introducing vertices with Kirchhoff coupling condition at the
distance $l_0 := \mathrm{min\,}\{l_1',\dots,l_p'\}$ from
$\mathcal{X}_1$, we can employ the model sketched in
Figure~\ref{fig4}. One defines the subspace of weighted symmetric
functions as
  % ------------- %
  $$
    g_\mathrm{sym1} (x) = \frac{1}{\sqrt{\sum_{j=1}^p c_j}}\sum_{j=1}^p
    \sqrt{c_j} g_j (x)\,,\quad g_\mathrm{sym2} (x) =
    \frac{1}{\sqrt{\sum_{j=1}^q \tilde c_j}}\sum_{j=1}^q
    \sqrt{\tilde c_j} \tilde g_j (x)\,,
  $$
  % ------------- %
where $g_j$'s and $\tilde g_j$'s refer to the internal and
external edges emanating from $\mathcal{X}_1$, respectively.
Following the argument of Section~\ref{reduction}, the restriction
of the Hamiltonian to this subspace can be unitarily transformed
to the Hamiltonian on the segment of length $l_0$ and the
half-line coupled mutually by the following condition
  % ------------- %
  $$
    \frac{g_\mathrm{sym1}(\mathcal{X}_1) }{\sqrt{\sum_{j=1}^p c_j}} =
    \frac{g_\mathrm{sym2}(\mathcal{X}_1) }{\sqrt{\sum_{j=1}^q \tilde c_j}} \,,
    \quad \sqrt{\sum_{j=1}^p c_j} \, g_\mathrm{sym1}'(\mathcal{X}_1) +
    \sqrt{\sum_{j=1}^q \tilde c_j}\, g_\mathrm{sym2}'(\mathcal{X}_1) = 0
  $$
  % ------------- %
at $\mathcal{X}_1$, where $g_\mathrm{sym1}(\mathcal{X}_1)$ and
$g_\mathrm{sym2}(\mathcal{X}_1)$ refer to the limit of the
functional values from the segment and the half-line,
respectively, and $g_{\mathrm{sym}1}'(\mathcal{X}_1)$,
$g_{\mathrm{sym}2}'(\mathcal{X}_1)$ stand for the appropriate
outward derivatives. Similarly, the restriction of the Hamiltonian
to the orthogonal complement of the subspace of symmetric
functions leads to the Dirichlet conditions from both sides of
$\mathcal{X}_1$. Consequently, the effective size of the graph is
less that its actual size if and only if $\sum_{j = 1}^p c_j =
\sum_{j = 1}^q \tilde c_j$ at some vertex.
  \end{proof}
  % ------------- %

The material in this section shows that within the category of
weighted graphs with appropriately generalized Kirchhoff boundary
conditions, the topology of the graph has no relevance to the
question of non-Weyl resonance asymptotics. After fixing the
topology of the graph, one can convert it to or from the non-Weyl
type by altering the weights on the leads in a suitable manner.

\section*{Acknowledgments}
We dedicate this paper to memory of Pierre Duclos, a good
colleague, and to the first author a friend and collaborator of
many years. The research was supported by the Czech Ministry of
Education, Youth and Sports within the project LC06002. We thank
Sasha Pushnitski for useful comments and the referees for comments
which helped to improve the article.

\end{document}